\documentclass[%
reprint,
superscriptaddress,
amsmath,amssymb,
prb,
nofootinbib,
]{revtex4-2}

\usepackage{subcaption}
\usepackage{graphicx}
\usepackage{bm}
\usepackage{hyperref}

\usepackage{xcolor}

\begin{document}

\title{Pressure-dependent melting and crystallization of B2-NiAl from neural-network molecular dynamics}
	
	\author{A. S. Onegin}
	\email{onegin.as@phystech.edu}
	\author{P. R. Levashov}
	\affiliation{%
	Joint Institute for High Temperatures, 
	Izhorskaya 13 Bldg 2, Moscow 125412, Russia
}
	\author{N. M. Chtchelkatchev}
	\affiliation{%
		Joint Institute for High Temperatures, 
		Izhorskaya 13 Bldg 2, Moscow 125412, Russia
}
	\affiliation{%
		Joint Institute for Nuclear Research, 6 Joliot-Curie Street, Dubna, 141980, Moscow Region, Russia
}

	\date{\today}
		
	\begin{abstract}
		We investigate pressure-dependent melting of ordered B2-NiAl using neural-network molecular dynamics with a Deep Potential interatomic model. Melting temperatures are determined from two-phase solid--liquid coexistence simulations over a broad pressure range, yielding the melting curve \(T_m(P)\). Relative to available experimental and previous molecular-dynamics results, the present calculations predict a stronger increase of the melting temperature with pressure at elevated compression. To assess the thermodynamic consistency of the calculated melting line, we evaluate the enthalpy and volume changes upon melting and compare the Clapeyron slope with the derivative of the fitted \(T_m(P)\) curve. The two estimates are in good agreement over most of the investigated pressure range, supporting the internal consistency of the coexistence results. To probe the character of melting, we perform a layer-resolved composition analysis of the coexistence configurations and find that the coexisting liquid remains essentially equiatomic at all studied pressures, with deviations of the aluminum fraction from the stoichiometric value not exceeding \(5\times10^{-3}\). This provides direct atomistic evidence that melting of B2-NiAl remains congruent within the present model. Together, these results establish a thermodynamically consistent pressure-dependent melting description of B2-NiAl and clarify the character of its melting under compression.

	\end{abstract}

	\maketitle
	
	
\section{\label{sec:introduction} Introduction}

Nickel aluminide NiAl is a prototypical ordered intermetallic compound with a high melting temperature, low density, good thermal conductivity, and oxidation resistance. These properties make Ni--Al intermetallics important for high-temperature structural applications, protective coatings, and reactive-material systems~\cite{morsi2001reaction,zhao2006atomistic,xie2023atomistic}. At the equiatomic composition, NiAl crystallizes in the ordered B2~(CsCl-type) structure, a chemically ordered bcc-derived arrangement
consisting of two interpenetrating sublattices preferentially occupied
by Al and Ni. This phase is central to the Al--Ni phase diagram and is a key reference for phase stability, ordering, and solidification in this binary system~\cite{okamoto1993ni,okamoto2004ni,goiri2016phase,davey2020first_CALPHAD}.

Despite the long-standing interest in NiAl, its melting behavior under pressure
is still not fully constrained. At ambient pressure, stoichiometric B2-NiAl is
commonly treated as a congruently melting compound. Under compression, however,
the melting temperature, volume and enthalpy changes, and possible
solid--liquid partitioning must be established explicitly. This is important
for atomistic modeling, in which short simulation time may suppress chemical
redistribution, and interatomic potentials may bias thermodynamic and ordering
properties. For Ni$_{50}$Al$_{50}$, atomistic simulations have shown that
chemical ordering and many-body interactions strongly affect nucleation and
crystal growth~\cite{tang2018chemical,tang2018composition,yang2021crystal}.

Molecular-dynamics simulations of the pressure-dependent melting of B2-NiAl
have so far mainly used embedded-atom-type potentials~\cite{purja2009development}.
Zhang \textit{et al.} calculated the melting curve using both one-phase and
two-phase approaches and fitted it by a Simon--Glatzel form~\cite{zhang2014molecular}.
That work provided an important reference for \(T_m(P)\). However, it predated
direct high-pressure melting measurements for NiAl and relied on the
transferability of the empirical potential. Other Ni--Al potentials have also
been proposed for reactive-alloying simulations~\cite{izvekov2012free}.
Recently, Bulatov \textit{et al.} reported experimental melting-line points for
nickel monoaluminide at elevated pressures~\cite{bulatov2025melting}.
Shock-compression experiments provide complementary high-pressure information
on NiAl~\cite{yakushev2019shock}. These data call for a reassessment of the
B2-NiAl melting curve using more accurate large-scale atomistic models.

Machine-learning interatomic potentials offer this possibility. Neural-network
and Deep Potential models can reproduce \textit{ab initio} potential-energy surfaces at
a cost suitable for large-scale molecular dynamics
~\cite{behler2007generalized,zhang2018deep,wang2018deepmd,mueller2020machine}.
Their accuracy depends on the representation of local atomic environments
~\cite{behler2011atom,onat2020sensitivity}. Modern implementations, including
DeePMD-kit v2 and GPUMD, make such models practical for large-scale simulations~\cite{zeng2023deepmd,fan2022gpumd}. For ordered binary alloys, the potential must describe both phase energetics and local chemical order. Machine-learned simulations of equiatomic Al--Ni melts have been recently used to study structure, transport, and B2 nucleation~\cite{sandberg2025homogeneous}.

Several methods exist to determine melting temperatures, including homogeneous heating, hysteresis protocols, interface pinning, free-energy calculations, and direct solid--liquid coexistence simulations~\cite{zou2020investigation}. Here we use the two-phase coexistence approach, which follows a solid--liquid interface directly and avoids most of the superheating typical of one-phase heating. This method has been successfully applied to pressure-dependent melting curves in molecular-dynamics simulations~\cite{ladd1978interfacial,morris2002melting,dozhdikov2012,yoo2004melting,alfe2002complementary,alfe2009temperature,vovcadlo2004ab,cazorla2007ab}. In addition to determining the melting temperature itself, the thermodynamic consistency of the resulting melting curve can be assessed through the Clapeyron relation, which connects its pressure derivative to the enthalpy and volume changes upon melting
~\cite{bonev2004quantum,aguado2005new,tangney2009melting,minakov2022ab, goswami2025high}. We therefore complement the two-phase coexistence
calculations with independent single-phase simulations of the solid and liquid at the calculated melting temperatures, allowing the Clapeyron slope to be compared directly with the derivative of the fitted melting curve.

A separate question is whether melting remains congruent under compression. In congruent melting, the coexisting solid and liquid have the same composition, whereas non-congruent melting implies chemical partitioning between the phases. For B2-NiAl this cannot be inferred from \(T_m(P)\) alone. The composition of the
liquid region must be measured directly in the two-phase cell. In addition,
geometric bcc-like order and chemical B2 order should be distinguished, because
local bcc coordination does not imply ordering of the Al and Ni sublattices.
This distinction is important for Ni$_{50}$Al$_{50}$, where chemical ordering
has been shown to affect crystallization and interface growth
~\cite{tang2018chemical,tang2018composition,yang2021crystal}.

Here we study pressure-dependent melting of B2-NiAl using neural-network
molecular dynamics with a Deep Potential interatomic model. The melting curve
is determined from two-phase solid--liquid coexistence simulations from ambient
pressure to tens of gigapascals and is compared with available experiment and
previous MD results. We test thermodynamic consistency by comparing the
Clapeyron slope with the derivative of the fitted melting curve. Finally, we
perform layer-resolved composition analysis of the two-phase trajectories and
structural analysis of quenched configurations to distinguish bcc-like geometry
from true B2 chemical order. This allows us to determine both \(T_m(P)\) and
the congruence of melting within the present atomistic model.

\section{\label{sec:computational_methods} Computational methods}

\subsection{\label{sec:dp_construction}
	Training-database construction and high-pressure assessment of the Deep Potential model}

The interatomic potential employed in the present work was trained on an
extended Al--Cu--Ni/Al--Ni database constructed to cover pressure ranges relevant to the present melting calculations. The starting point was the first-principles database previously developed
for liquid Al--Cu--Ni alloys~\cite{ryltsev2022deep}, containing 30\,424
configurations with energies, atomic forces, and virial tensors calculated
using density functional theory~(DFT). For the
present study, this database was substantially extended by including ordered Al--Ni structures available in the Materials Project
\cite{jain2013commentary}, together with distorted crystalline, molten, and surface configurations.

To extend the database toward compressed and high-temperature states,
a large pool of candidate configurations was generated by molecular
dynamics using PET-MAD-S~\cite{mazitov2025pet}, the small-size variant
of the PET-MAD universal machine-learning interatomic potential based
on the Point Edge Transformer (PET) architecture and trained on the
Massive Atomistic Diversity (MAD) dataset. PET-MAD-S was used only as
a configuration generator; none of its predicted energies, forces, or
stresses were retained as reference data. The sampling covered pressures from 0 to 100~GPa in increments of 5~GPa and temperatures
from 300~K to approximately 1000~K above the largest estimated melting
temperature at the corresponding pressure. This produced approximately
$10^6$ candidate configurations.

Because consecutive MD snapshots are strongly correlated, the candidate pool was reduced before DFT labeling. Each configuration was represented by an averaged and normalized Smooth Overlap of Atomic Positions descriptor (SOAP)~\cite{bartok2013representing}, using a cutoff radius of 4.2~\AA, four radial basis functions, $l_{\max}=3$, and a Gaussian width of 0.52~\AA. A structurally diverse subset of approximately 1000 configurations was then selected by farthest-point sampling (FPS)~\cite{imbalzano2018automatic} and recalculated with DFT.

The database was subsequently refined through iterative cross-architecture active learning following the strategy of
Ref.~\cite{chtchelkatchev2026adaptive}. Two models with substantially
different representations were trained on the same evolving database: a
Deep Potential model based on the DeepPot-SE \texttt{se\_e2\_a}
descriptor~\cite{zhang2018end,wang2018deepmd}, trained from random
initialization, and the pretrained DPA-3.1-3M model
\cite{zhang2026graph}, which was fine-tuned on the present data. Both models were used to evaluate configurations generated during MD exploration, and their disagreement in energy per atom, atomic forces, stress tensors, and the hydrostatic-pressure contribution was used to identify candidate configurations for additional DFT labeling. The flagged structures were again reduced by SOAP/FPS to avoid redundant sampling, labeled by DFT, and added to the database. The train--sample--select--label cycle was continued until no substantial new population of high-disagreement configurations was generated.

Bulk sampling was additionally supplemented by zero-pressure slab
configurations constructed from Al--Ni crystalline phases and sampled from 300~K to their estimated melting temperatures. These configurations provided undercoordinated, surface-relaxed, and near-surface liquid environments. Altogether, the enrichment procedure added approximately 6800 DFT-labeled configurations, resulting in a final database of about $3.72\times10^4$ configurations. The database therefore contains explicitly sampled compressed configurations over a pressure interval substantially wider than the 0--50~GPa range used for the melting calculations.

The production Deep Potential model employed the \texttt{se\_e2\_a}
descriptor with a cutoff radius of 7.0~\AA\ and a smoothing radius of
1.3~\AA. The descriptor network contained three hidden layers with
24, 48, and 96 neurons and an axis-neuron size of 16, while the fitting
network contained three hidden layers of 240 neurons. The neighbor-selection parameters for Al, Cu, and Ni were 160, 136, and 196, respectively. Hyperbolic-tangent activation functions and double precision were used. The model was trained for $2\times10^6$ steps with a batch size of 4 using the Adam optimizer. The learning rate was exponentially decreased from $10^{-3}$ to $3.51\times10^{-8}$ with a decay interval of 5000 steps. During training, the energy and virial prefactors were increased from 0.02 to 1, whereas the force prefactor was decreased from 1000 to 1.

The additional DFT labeling calculations were performed as static
single-point calculations in VASP~\cite{hafner2008ab}, using the same
electronic-structure convention as for the seed database. DFT energies,
atomic forces, and stress tensors were retained as reference quantities for training.

\begin{figure}[h]
	\centering
	\includegraphics[width=0.98\columnwidth]
	{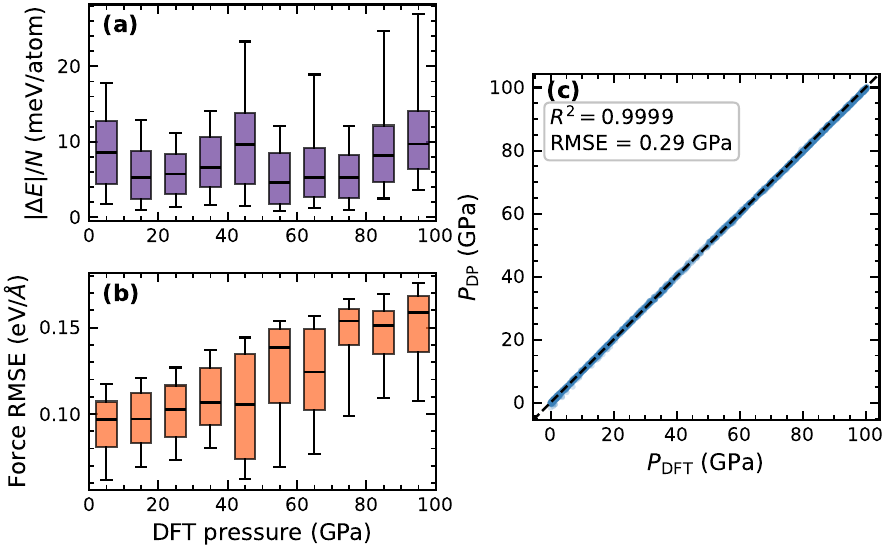}
	\caption{\label{fig:val}
		Pressure-dependent accuracy of the Deep Potential model with respect to
		the DFT reference data.
		(a) Absolute energy error per atom and
		(b) force RMSE grouped into 10~GPa intervals of the DFT hydrostatic
		pressure. Boxes indicate the interquartile range, horizontal lines denote
		the median, and whiskers correspond to the 10th and 90th percentiles.
		(c) Hydrostatic pressure predicted by the Deep Potential model versus the
		DFT reference pressure. The dashed line denotes
		$P_{\mathrm{DP}}=P_{\mathrm{DFT}}$. The pressure predictions yield
		$R^2=0.9999$ and an RMSE of 0.29~GPa.
	}
\end{figure}

The pressure dependence of the model errors is summarized in
Fig.~\ref{fig:val}. The absolute energy errors and force root-mean-square error~(RMSE) values were grouped into 10~GPa intervals according to the DFT hydrostatic pressure. The typical energy error remains of the order of 10~meV/atom, while the force RMSE remains below approximately 0.16~eV/\AA, without a pronounced increase within the 0--50~GPa interval used for the melting simulations. The predicted hydrostatic pressure was additionally compared directly with the DFT value obtained from the stress tensor, yielding $R^2=0.9999$ and an RMSE of 0.29~GPa. Although the absolute errors increase moderately with pressure, particularly
for the atomic forces, their magnitude remains limited over the entire
pressure interval considered; in particular, the force RMSE remains below
approximately 0.16~eV/\AA.

\subsection{\label{sec:model}Deep Potential model and molecular-dynamics setup}

All molecular-dynamics simulations in the present work were performed with
the LAMMPS package~\cite{LAMMPS} using a machine-learning interatomic potential of the Deep Potential type implemented in the DeepMD-kit~\cite{wang2018deepmd} framework. The potential was used to describe interatomic interactions in the Al--Ni system over the entire pressure and temperature range considered in this study.

Unless stated otherwise, the simulations were carried out for supercells
containing \(N=16384\) atoms. Periodic boundary conditions were applied in
all three spatial directions. The equations of motion were integrated with a
time step of 4~fs. Depending on the specific stage of the calculation,
isothermal--isobaric (\(NPT\)), constant-pressure--constant-enthalpy
(\(NPH\)), and canonical (\(NVT\)) ensembles were employed.

\subsection{Two-phase coexistence protocol for $T_m$($P$)}

The melting temperature was determined using the two-phase coexistence
method. This approach follows the general idea of the solid--liquid
coexistence technique proposed by Ladd and Woodcock~\cite{Ladd:CPL:1977}
and further discussed in comparative studies of melting-point calculation methods~\cite{dozhdikov2012,zou2020investigation}. In this method, solid and liquid regions are prepared within the same periodic simulation cell, and the system is subsequently allowed to evolve at a prescribed pressure. If the total enthalpy corresponds to the coexistence region, the solid--liquid interface remains stable, and the average temperature of the final trajectory provides an estimate of the melting temperature.

The initial configuration was a crystalline supercell of the corresponding phase. The system was first equilibrated in the isothermal--isobaric ensemble at the target pressure $P$ and an estimated melting temperature $T_{\mathrm{es}}$. During this stage, the
temperature was gradually increased from 300~K to $T_{\mathrm{es}}$, while the pressure was controlled isotropically,
\begin{equation}
	P_{xx} = P_{yy} = P_{zz} = P .
\end{equation}
This preliminary stage was used to obtain a relaxed crystalline configuration at the required pressure.

The B2 crystal was oriented with the cubic [100], [010], and [001] directions along the $x$, $y$, and $z$ axes of the simulation cell, respectively. The simulation cell was then divided into two parts along the $z$ direction. Consequently, the solid--liquid interfaces were parallel to the (001) crystallographic planes, which belong to the $\{100\}$ family of symmetry-equivalent planes in cubic B2-NiAl.
The dividing plane was chosen as
\begin{equation}
	z_{\mathrm{mid}} =
	\frac{z_{\mathrm{lo}} + z_{\mathrm{hi}}}{2}.
\end{equation}
Atoms with $z < z_{\mathrm{mid}}$ were assigned to the solid region, whereas atoms with $z > z_{\mathrm{mid}}$ were assigned to the region to be melted. The atoms in the solid part were temporarily fixed by setting their forces and velocities to zero. The other half of the system was heated in the canonical ensemble to a high temperature $T_{\mathrm{h}}$, chosen well above the expected melting temperature,
\begin{equation}
	T_{\mathrm{h}} > T_{\mathrm{m}} .
\end{equation}
As a result, a two-phase configuration containing crystalline and liquid
regions was formed. Due to periodic boundary conditions, two solid--liquid interfaces were present in the simulation cell.

After the liquid region had been created, the constraints imposed on the solid part were removed. The entire two-phase system was then relaxed at the target pressure and at $T_{\mathrm{es}}$ in the $NPT$ ensemble. This step was used to remove artificial stresses introduced during the construction of the solid--liquid configuration and to stabilize the interfaces before the production coexistence run. The two-phase configuration was relaxed in the $NPT$ ensemble for 40\,000 time steps, corresponding to 160~ps.

The final stage was performed in the constant-pressure,
constant-enthalpy (\(NPH\)) ensemble. The production coexistence trajectory was run for 300\,000 time steps,
corresponding to 1.2~ns at the 4~fs integration time step. At this stage, no thermostat was applied; therefore, the instantaneous temperature was not imposed externally but was determined by the enthalpy and the relative amounts of the solid and liquid phases. If $T_{\mathrm{es}}$ was below the melting temperature, the liquid part crystallized during the simulation. If $T_{\mathrm{es}}$ was above the melting temperature, the crystalline part melted. The value of $T_{\mathrm{es}}$ was therefore varied until a stable solid--liquid coexistence state was observed over the final part of the
trajectory.

The melting temperature at a given pressure was determined as the average
instantaneous temperature over the stationary part of the \(NPH\) trajectory,
after the initial relaxation of the solid--liquid interface. In practice, the
averaging interval was selected from the portion of the trajectory where a
stable coexistence state had already been established and no systematic drift
of the temperature was observed. The statistical uncertainty of the melting temperature was estimated by block averaging. For this purpose, the stationary segment of the trajectory was divided into several consecutive blocks of equal length, and the average temperature was calculated for each block. The uncertainty of the final melting temperature was then taken as the standard deviation of these
block-averaged temperatures.

\subsection{\label{sec:clapeyron_method}Single-phase solid/liquid calculations of $\Delta H$, $\Delta V$, and Clapeyron slope}

To characterize the thermodynamic signature of melting and to provide an
independent consistency check for the calculated melting curve, we performed additional single-phase simulations of the solid and liquid states at the melting temperature $T_m(P)$ obtained from the two-phase coexistence calculations.

For each pressure, two separate simulations were carried out: one for the
crystalline B2 phase and one for the liquid phase. The crystalline system was
equilibrated directly in the isothermal--isobaric ensemble at the target
pressure $P$ and temperature $T_m(P)$. The liquid system was prepared by
heating the crystal to a temperature approximately 2000~K above the
corresponding melting temperature. The formation of a liquid state was
additionally verified by visual inspection of the atomic configurations, by
checking the spatial homogeneity of the layer-resolved density/composition
profiles, and from the radial distribution function, which showed the loss of
long-range crystalline order. The liquid was then cooled to $T_m(P)$ and
equilibrated in the same $NPT$ ensemble at pressure $P$. After equilibration,
production runs were performed and the instantaneous enthalpy and volume were
sampled along the trajectories.

The average enthalpy and volume per atom for phase $\alpha$ were computed as
\begin{equation}
	h_\alpha(P,T_m)
	=
	\frac{\langle E_\alpha \rangle
		+ P\langle V_\alpha \rangle}{N},
	\quad
	v_\alpha(P,T_m)
	=
	\frac{\langle V_\alpha \rangle}{N},
	\label{eq:hv_per_atom}
\end{equation}
where $\alpha=\mathrm{sol},\mathrm{liq}$, $N$ is the number of atoms,
$\langle E_\alpha \rangle$ and $\langle V_\alpha \rangle$ are the
time-averaged total energy and simulation-cell volume, respectively, and
$P$ is the prescribed pressure of the $NPT$ ensemble. Thus, the enthalpy
was evaluated using the target ensemble pressure rather than the
instantaneous pressure.

The enthalpy and volume changes upon melting were then defined as
\begin{equation}
	\Delta h = h_{\mathrm{liq}} - h_{\mathrm{sol}},
	\qquad
	\Delta v = v_{\mathrm{liq}} - v_{\mathrm{sol}}.
	\label{eq:delta_hv}
\end{equation}

The local slope of the melting curve was estimated from the Clapeyron relation,
\begin{equation}
	\frac{dT_m}{dP}
	=
	\frac{T_m \Delta v}{\Delta h},
	\label{eq:clapeyron}
\end{equation}
which links the slope of the coexistence line to the latent enthalpy and volume
change of melting.

In addition to the Clapeyron estimate, the slope of the melting curve was evaluated from the analytical derivative of the Simon--Glatzel fit to the calculated melting temperatures,
\begin{equation}
	T_m(P) = T_0 \left( 1 + \frac{P}{a} \right)^c,
	\label{eq:sg_fit}
\end{equation}
which gives
\begin{equation}
	\frac{dT_m}{dP}
	=
	\frac{T_0 c}{a}
	\left( 1 + \frac{P}{a} \right)^{c-1}.
	\label{eq:sg_derivative}
\end{equation}
The uncertainty band of the fitted derivative shown in
Fig.~\ref{fig:comparison_dtmdp} reflects the uncertainty of the fit
parameters.

\subsection{\label{sec:composition_method}Layer-resolved composition analysis}

To analyze the local chemical composition in the solid--liquid coexistence simulations, we constructed layer-resolved profiles along the [001] direction. For each trajectory frame, the atomic
coordinates were transformed to the reduced coordinate
\begin{equation}
	s = \frac{z}{L_z},
\end{equation}
where \(L_z\) is the instantaneous simulation-cell length along the
\(z\)-axis. The interval \(0 \le s < 1\) was divided into 200 uniform bins.

Before temporal averaging, the layer-resolved profiles were aligned to account for the translational displacement of the two-phase density pattern within the periodic simulation cell. A reference total-density profile was constructed from the first analyzed frame. For each subsequent frame, the total-density profile along \(s\) was evaluated, smoothed with a Gaussian filter, and compared with the reference profile. The optimal translational shift was determined from the maximum of the cyclic cross-correlation function between the two profiles. This shift was applied only during post-processing, ensuring that the solid, interfacial, and liquid regions remained spatially aligned when the composition profiles were averaged.

After alignment, the numbers of Al and Ni atoms in each bin,
\(N_{\mathrm{Al}}(s)\) and \(N_{\mathrm{Ni}}(s)\), were accumulated over all analyzed frames. The layer-resolved composition profiles were then computed as
\begin{equation}
	x_{\mathrm{Al}}(s)
	=
	\frac{N_{\mathrm{Al}}(s)}
	{N_{\mathrm{Al}}(s)+N_{\mathrm{Ni}}(s)}.
\end{equation}
The final profiles were obtained by averaging over all processed frames. The statistical uncertainty of the liquid composition was estimated by block averaging of the stationary part of the coexistence trajectory.
The trajectory was divided into consecutive blocks of equal length, and
the average liquid-region composition was evaluated independently for
each block. The standard error of the mean (SEM) was then calculated
from the resulting block averages. These uncertainties are shown as
error bars in Fig.~\ref{fig:delta_x}.

\subsection{\label{sec:ptm_method}PTM and B2-order analysis of quenched configurations}

To characterize the structural state of the undercooled liquid after
quenching, the atomic configurations were analyzed using the
polyhedral template matching (PTM)~\cite{larsen2016robust} method as implemented in OVITO~\cite{ovito}.
This analysis was performed for the final configurations obtained after
quenching at different cooling rates and at selected pressures.

The PTM algorithm classifies each atom according to the local geometry of
its nearest-neighbor environment. In the present work, atoms were assigned to
the mutually exclusive body-centered cubic (BCC), face-centered cubic (FCC),
hexagonal close-packed (HCP), icosahedral (ICO), or \textit{Other} categories
using a root-mean-square deviation~(RMSD) cutoff of 0.1; the corresponding fractions therefore sum to
100\%. The \textit{Other} category includes atoms whose local environments do
not match any of the considered structural templates within this cutoff.

The B2 structure of equiatomic NiAl is a chemically ordered bcc-derived
structure, in which Al and Ni preferentially occupy two interpenetrating
sublattices. Geometrically, its local environment is therefore identified by
PTM as BCC-like. Since PTM is purely geometrical, however, it cannot
distinguish chemically ordered B2 environments from chemically disordered
bcc-like ones. Thus, B2 is not treated as an independent PTM structure type,
but rather as a chemically ordered subset of the atoms classified as
BCC-like.

To quantify this chemical ordering, an additional local B2 order parameter
was introduced. For each atom \(i\), the eight nearest neighbors
corresponding to the first coordination shell of the B2 structure were
identified, and the local chemical order parameter was defined as
\begin{equation}
	q_i^{\mathrm{B2}} =
	\frac{N_{\mathrm{opp}}^{(1)}}{8},
	\label{eq:qB2}
\end{equation}
where \(N_{\mathrm{opp}}^{(1)}\) is the number of nearest neighbors of the
opposite chemical species. Thus, for an Al atom, \(N_{\mathrm{opp}}^{(1)}\)
counts the number of Ni atoms among its eight nearest neighbors, while for
a Ni atom it counts the number of Al neighbors. In an ideal B2 crystal,
\(q_i^{\mathrm{B2}}=1\), whereas for a chemically disordered 50:50
arrangement the expected value is close to \(0.5\).

An atom was classified as \emph{B2-like} if two conditions were satisfied
simultaneously: (i) its local geometry was identified by PTM as BCC-like,
and (ii) its local chemical order parameter exceeded a prescribed threshold,
\begin{equation}
	q_i^{\mathrm{B2}} \ge q_{\mathrm{cut}}.
\end{equation}
In the present analysis, the threshold \(q_{\mathrm{cut}}=0.875\) was used,
which corresponds to at least seven out of eight nearest neighbors having
the chemically correct type. This choice provides a reasonably strict
criterion for near-B2 local ordering while remaining tolerant to thermal
fluctuations and local defects.

For each analyzed configuration, three quantities were monitored:
\begin{enumerate}
	\item the total fraction of BCC-like atoms,
	\begin{equation}
		f_{\mathrm{BCC}} = \frac{N_{\mathrm{BCC}}}{N},
	\end{equation}
	where \(N_{\mathrm{BCC}}\) is the number of atoms classified by PTM as
	BCC-like and \(N\) is the total number of atoms;
	
	\item the fraction of B2-like atoms,
	\begin{equation}
		f_{\mathrm{B2}} =
		\frac{N_{\mathrm{BCC}\,\cap\,q_i^{\mathrm{B2}}\ge q_{\mathrm{cut}}}}{N},
	\end{equation}
	which measures the fraction of atoms that are both geometrically
	BCC-like and chemically close to the B2 arrangement;
	
	\item the mean B2 order parameter evaluated over the subset of BCC-like
	atoms,
	\begin{equation}
		\left\langle q^{\mathrm{B2}} \right\rangle_{\mathrm{BCC}}
		=
		\frac{1}{N_{\mathrm{BCC}}}
		\sum_{i \in \mathrm{BCC}} q_i^{\mathrm{B2}}.
		\label{eq:qB2_bcc_mean}
	\end{equation}
\end{enumerate}

The temperature evolution of these quantities during quenching was obtained
by combining the frame-resolved OVITO analysis with the temperature
recorded in the molecular-dynamics trajectories. The resulting dependences
\(f_{\mathrm{BCC}}(T)\), \(f_{\mathrm{B2}}(T)\), and
\(\left\langle q^{\mathrm{B2}} \right\rangle_{\mathrm{BCC}}(T)\) were used
to distinguish between the formation of a bcc-like geometric framework and
the subsequent development of chemical B2 ordering.

The present PTM analysis should be distinguished from the cluster-alignment
approach used, for example, by Yang \textit{et al.} to analyze short-range
order in Pd--Si liquids~\cite{yang2019effects}. Both approaches compare
local atomic environments with reference motifs, but they serve somewhat
different purposes. The cluster-alignment method directly optimizes the
geometrical alignment of first-neighbor clusters with a selected library
of crystalline and non-crystalline motifs and is particularly useful for
identifying glass-forming short-range-order motifs. In the present work,
PTM is used to provide a robust classification of the local geometrical
environment into BCC, FCC, HCP, ICO, or \textit{Other} categories.
Chemical B2 ordering is then quantified independently using the local
order parameter defined below.

\section{\label{sec:melting_curve} Results and Discussion}

\subsection{Pressure-dependent melting curve of B2-NiAl}

\begin{figure}[h]
	\centering
	\includegraphics[width=0.95\columnwidth]{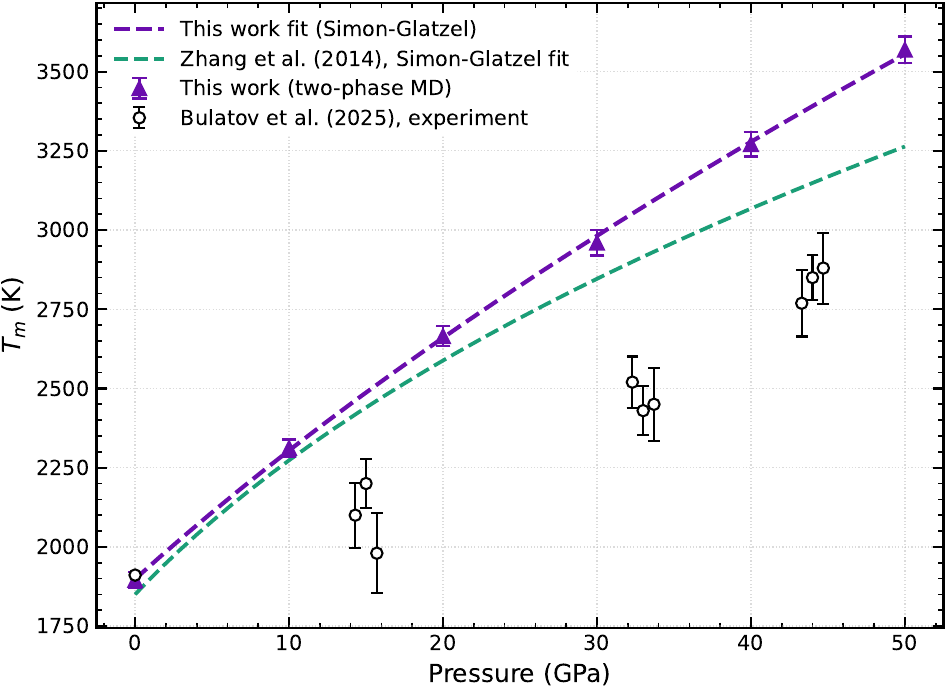}
	\caption{\label{fig:tm_vs_p}
		Melting curve of B2-NiAl as a function of pressure.
		Filled triangles show the present two-phase MD results with vertical
		error bars. The dashed purple line is the Simon--Glatzel fit to the
		present data. Open circles show the experimental data of Bulatov
		\textit{et al.}~\cite{bulatov2025melting}; the scatter of the points at each
		pressure corresponds to separate heating series reported in that work.
		The green dashed line shows the Simon--Glatzel fit from Zhang
		\textit{et al.}~\cite{zhang2014molecular}. The present melting curve agrees
		closely with the ambient-pressure experimental melting point and is
		close to the EAM-based curve up to about 20~GPa, but predicts a
		stronger pressure dependence at higher pressures.
	}
\end{figure}

The pressure-dependent melting temperatures of B2-NiAl obtained in the present
two-phase coexistence simulations are listed in
Table~\ref{tab:melting_curve_this_work} and shown in Fig.~\ref{fig:tm_vs_p}.
The calculations cover the pressure range from ambient pressure to 50~GPa.
The melting temperature increases monotonically from
$1896\pm 9$~K at 0~GPa to $3569\pm 17$~K at 50~GPa.

\begin{table}[h]
	\centering
	\caption{Melting curve of B2-NiAl calculated in this work using two-phase
		MD simulations.}
	\label{tab:melting_curve_this_work}
	\begin{tabular}{cc}
		\hline
		Pressure (GPa) & $T_m$ (K) \\
		\hline
		0  & $1896 \pm 9$ \\
		10 & $2312 \pm 13$ \\
		20 & $2665 \pm 14$ \\
		30 & $2960 \pm 18$ \\
		40 & $3271 \pm 17$ \\
		50 & $3569 \pm 17$ \\
		\hline
	\end{tabular}
\end{table}

At ambient pressure, the present value is very close to the experimental
melting temperature of stoichiometric NiAl,
$T_m^{\rm exp}=1911$~K, reported in the equilibrium phase-diagram
compilation~\cite{massalski1986binary} and used as the reference value in previous molecular-dynamics studies~\cite{zhang2014molecular}. The deviation of the present result
from this value is only about 15~K, or approximately 0.8\%. This agreement is
important because the zero-pressure melting point provides a direct test of
the quality of the interatomic model in the temperature range relevant to
melting.

As an additional check of the DeepMD model in limiting elemental systems, we
also calculated the melting temperatures of pure fcc Al and fcc Ni using the
same two-phase coexistence protocol. The resulting values are
$927\pm 5$~K for Al and $1726\pm 9$~K for Ni, 
respectively. They are
in close agreement with the experimental melting temperatures of Al and Ni,
approximately 933~K and 1728~K. This provides an additional validation of the
potential for both elemental endpoints of the Al--Ni system.

The present melting curve is compared with two available references. The first
one is the molecular-dynamics study of Zhang \textit{et al.}, who calculated the
melting curve of B2-NiAl using the embedded-atom-method potential of
Pun and Mishin~\cite{zhang2014molecular,purja2009development}. Their two-phase result was
represented by the Simon--Glatzel form (K, GPa)
\begin{equation}
	T_m(P) = 1850 	\left(
	1+\frac{P }{12.806}
	\right)^{0.357}.
	\label{eq:sg_zhang}
\end{equation}
The second reference is the recent experimental work of Bulatov
\textit{et al.}, where the onset of melting of nickel monoaluminide was
measured in a laser-heated diamond-anvil cell at 15, 33, and 44~GPa using
the dynamics of speckle patterns~\cite{bulatov2025melting}.

The comparison with the EAM-based results requires some caution. The
Pun--Mishin potential was not fitted directly to melting temperatures or to
liquid-state configurations. It was constructed within the EAM formalism from
previously developed elemental Ni and Al potentials, while the Ni--Al
cross-interaction was fitted mainly to the experimental cohesive energy,
lattice parameter and elastic constants of B2-NiAl, and to \textit{ab initio}
formation energies of several ordered compounds with different structures and
compositions~\cite{purja2009development}. Therefore, the melting behavior predicted
by this potential should be regarded as a transferability test rather than as a
fitted property. In the original validation by Pun and Mishin, the calculated
melting temperatures were 1701~K for Ni, 1678~K for Ni$_3$Al, 1780~K for
NiAl, and 1042~K for Al, compared with experimental values of 1728, 1645,
1911, and 933~K, respectively~\cite{purja2009development}. Thus, the EAM potential
reproduces Ni and Ni$_3$Al rather well, but shows larger deviations for NiAl
and especially for Al.

The experimental data of Bulatov \textit{et al.} also have to be interpreted
with care. Their method detects the onset of local melting during pulsed laser
heating in a diamond-anvil cell. Such measurements are affected by the strong
spatial heterogeneity of the temperature field in the heated region and by the
difficulty of identifying the beginning of partial melting. This is reflected in
the scatter of the experimental points obtained in separate heating series at
the same nominal pressures~\cite{bulatov2025melting}. Moreover, Bulatov
\textit{et al.} noted that a single Simon--Glatzel fit to all their data is not
fully satisfactory because it gives a downward-convex curve and poorly
constrained fitting coefficients. They therefore discussed a possible kink of
the melting line associated with a polymorphic transformation below the
melting curve~\cite{bulatov2025melting}.

As seen in Fig.~\ref{fig:tm_vs_p}, the experimental points of Bulatov
\textit{et al.} lie below both theoretical melting curves. The Simon--Glatzel
fit of Zhang \textit{et al.} lies above the experimental points, whereas the
present DeepMD fit lies still higher at elevated pressures. Up to about
20~GPa, the present results and the EAM-based curve of Zhang \textit{et al.}
are relatively close. At higher pressures, however, the difference becomes
substantial: in the range from 30 to 50~GPa, the present calculations predict
a noticeably steeper increase of $T_m$ with pressure. This stronger pressure
dependence suggests a different description of the relative thermodynamic
stability of compressed B2 solid and liquid NiAl by the present DeepMD model
and by the earlier EAM potential. This point is analyzed further below through
the volume and enthalpy changes on melting and the Clapeyron consistency
check.

For the present two-phase MD data, the same functional form was used. Thus, the
present melting curve can be written explicitly as
\begin{equation}
	T_m(P) = 1898.45 \left(
	1+\frac{P}{24.19}
	\right)^{0.5598}.
	\label{eq:sg_this_work}
\end{equation}
Here, $T_m$ is expressed in K and $P$ in GPa. The quality of the fit is
characterized by $R^2=0.99987852$ and a reduced chi-square of
$\chi^2_{\nu}=0.186732$.

\subsection{\label{sec:clapeyron} Thermodynamic signature of melting: $\Delta V$, $\Delta H$, and Clapeyron consistency}

\begin{figure}[h]
	\centering
	\includegraphics[width=0.95\columnwidth]{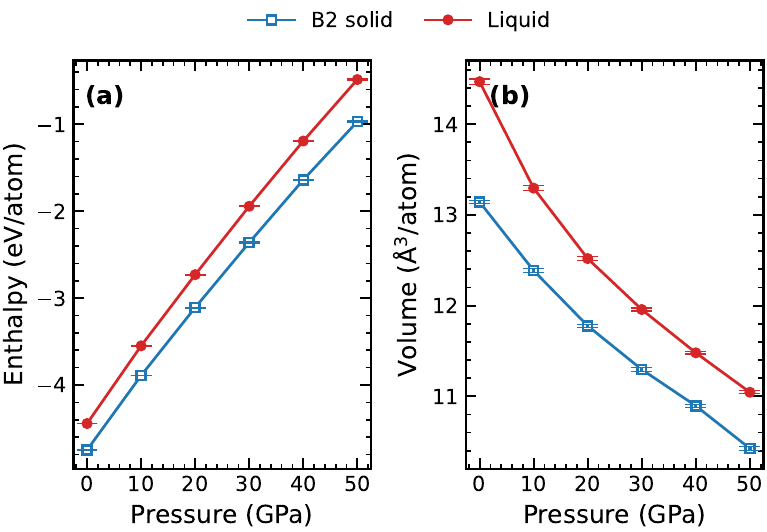}
	\caption{\label{fig:h_v}
		Pressure dependences of the enthalpy and volume per atom for the
		solid and liquid phases evaluated at the melting temperature.
		At all investigated pressures, the liquid phase has both a higher
		enthalpy and a larger atomic volume than the corresponding solid phase.
	}
\end{figure}

Figure~\ref{fig:h_v} shows the pressure dependences of the enthalpy and volume
per atom for the solid and liquid phases evaluated at the melting temperature.
At all investigated pressures, the liquid phase has a higher enthalpy and a
larger volume than the corresponding solid phase, as expected for normal
melting. Both the solid and liquid enthalpies increase monotonically with
pressure, while the atomic volumes decrease due to compression. The separation
between the solid and liquid branches remains clearly resolved over the entire
pressure range.

\begin{figure}[h]
	\centering
	\includegraphics[width=0.8\columnwidth]{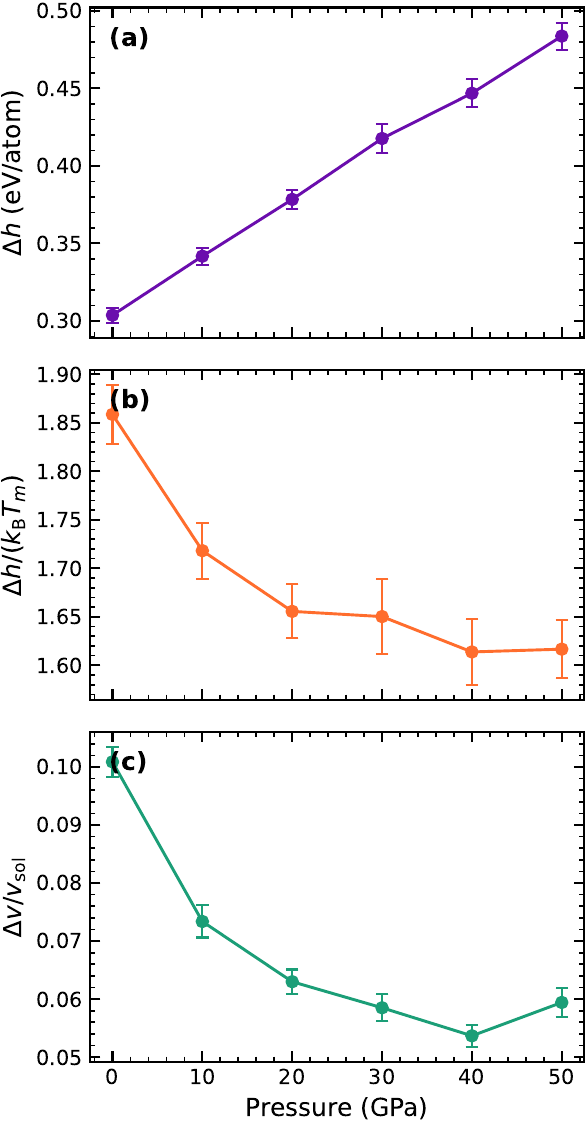}
	\caption{\label{fig:dh_d_v}
		Pressure dependences of the enthalpy and volume changes upon melting,
		\(\Delta h = h_{\mathrm{liq}} - h_{\mathrm{sol}}\) and
		\(\Delta v = v_{\mathrm{liq}} - v_{\mathrm{sol}}\), per atom.
		The enthalpy change increases with pressure, whereas the volume change
		generally decreases, indicating a reduced density contrast between the
		coexisting phases under compression.
	}
\end{figure}

The corresponding thermodynamic changes upon melting are presented in
Fig.~\ref{fig:dh_d_v}. The enthalpy change, \(\Delta h = h_{\mathrm{liq}} - h_{\mathrm{sol}}\), increases with pressure, indicating an increase of the latent enthalpy of melting under compression. However, when normalized by the thermal energy scale, the dimensionless quantity \(\Delta h/(k_{\mathrm{B}}T_m)\) exhibits only a weak decrease with pressure. Thus, the increase of the absolute \(\Delta h\) largely follows the increase of the melting temperature along the coexistence curve. The fractional volume change, \(\Delta v/v_{\mathrm{sol}}\), where \(\Delta v = v_{\mathrm{liq}} - v_{\mathrm{sol}}\), decreases more pronouncedly with pressure. This indicates that the relative volume, and hence density, contrast between the coexisting solid and liquid phases becomes progressively smaller under compression. A slight increase of \(\Delta v/v_{\mathrm{sol}}\) is observed at 50~GPa relative to 40~GPa; however, within the statistical uncertainty, it does not reveal a reversal of the overall decreasing trend.

\begin{figure}[h!]
	\centering
	\includegraphics[width=0.95\columnwidth]{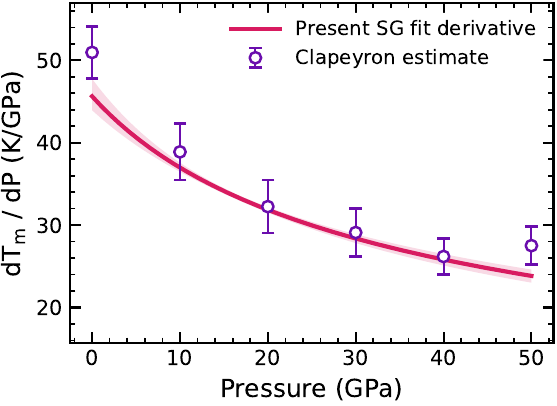}
	\caption{\label{fig:comparison_dtmdp}
		Comparison between the slope of the melting curve obtained from the
		Clapeyron relation and the derivative of the fitted melting curve,
		\(dT_m/dP\). The pink shaded band denotes the uncertainty of the fitted
		derivative. The agreement is particularly close in the intermediate
		pressure range, while at 0 and 50~GPa the Clapeyron estimates remain
		consistent with the fitted derivative within the error bars.
	}
\end{figure}

The pressure derivative of the melting temperature obtained from the
Clapeyron relation is compared in Fig.~\ref{fig:comparison_dtmdp} with the
derivative of the fitted melting curve, \(dT_m/dP\). Overall, the agreement is
very good. In the intermediate pressure range from 10 to 40~GPa, the
Clapeyron estimates lie essentially on the derivative of the fitted
\(T_m(P)\) curve. At the boundary points, 0 and 50~GPa, the agreement is less
exact, but the error bars of the Clapeyron estimates still overlap with the
uncertainty band of the fitted derivative. The pink shaded region in
Fig.~\ref{fig:comparison_dtmdp} indicates the uncertainty associated with the
fit of the melting curve. The consistency between the independently evaluated
Clapeyron slope and the derivative of the fitted melting line confirms that the
calculated melting curve is thermodynamically self-consistent within the
statistical uncertainty of the present simulations.

Taken together, these results show that melting of B2-NiAl in the investigated
pressure range is characterized by positive enthalpy and volume changes, a
progressively increasing latent enthalpy of melting, and a decreasing volume
contrast between the solid and liquid phases. This combination naturally leads
to a positive melting-curve slope, whose magnitude is well reproduced by the
Clapeyron relation.

\subsection{\label{sec:congruent_evidence}Direct atomistic evidence of congruent melting}

To test whether melting of B2-NiAl remains congruent under pressure, we
analyzed the chemical composition of the coexisting liquid phase in the
solid--liquid coexistence simulations. The key idea of this analysis is that,
for congruent melting, the liquid coexisting with the stoichiometric B2 solid
should also remain stoichiometric within statistical uncertainty.

\begin{figure}[h]
	\centering
	\includegraphics[width=0.9\columnwidth]{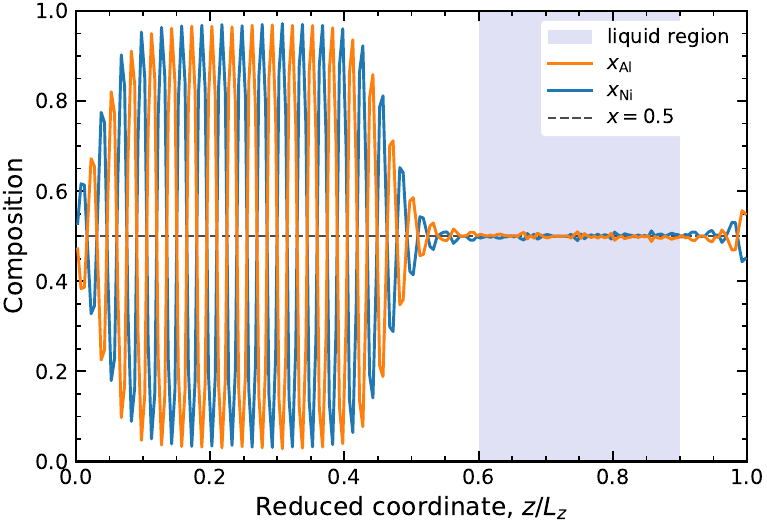}
	\caption{\label{fig:profile}
		Layer-resolved aluminum and nickel composition profiles along the
		$z\parallel[001]$ direction for the 0~GPa coexistence simulation. The left part of the cell corresponds
		to the ordered B2 solid, the middle part to the interfacial region, and the
		right part to the liquid phase. The shaded interval
		$0.6 \le z/L_z \le 0.9$ marks the bulk-liquid region used to evaluate the
		average liquid composition.
	}
\end{figure}

Figure~\ref{fig:profile} shows the layer-resolved composition profile for the 0~GPa coexistence simulation. In the left part of the cell, corresponding to the solid region, the aluminum and nickel fractions exhibit pronounced oscillations. This behavior reflects the chemical ordering of the B2 phase, where adjacent atomic planes along the $z\parallel[001]$ direction are alternately enriched in Al and Ni. Moving along the [001] direction one enters the interfacial region, after which the profile approaches a nearly constant value in the liquid part of the cell. The interval $0.6 \le z/L_z \le 0.9$, highlighted in the figure, was selected as the bulk-liquid region and used to evaluate the average liquid composition. Averaging the aluminum fraction over the selected liquid window yields a value close to 0.5, indicating that the coexisting liquid remains essentially stoichiometric.

\begin{figure}[h]
	\centering
	\includegraphics[width=0.95\columnwidth]{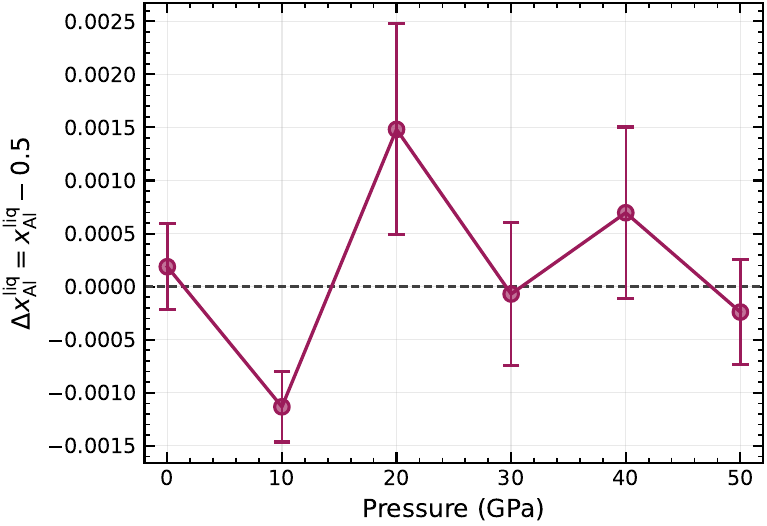}
	\caption{\label{fig:delta_x}
		Deviation of the aluminum fraction in the coexisting liquid phase
		from the stoichiometric value,
		$\Delta x_{\mathrm{Al}}^{\mathrm{liq}}
		= x_{\mathrm{Al}}^{\mathrm{liq}} - 0.5$, as a function of pressure.
		Error bars denote the standard error of the mean obtained by block
		averaging of the stationary coexistence trajectory. For all
		investigated pressures, the deviation remains below 0.005,
		indicating that the liquid phase stays essentially equiatomic within
		the statistical uncertainty of the simulations.
	}
\end{figure}

The same analysis was repeated for all investigated pressures. The resulting
deviation of the aluminum fraction in the liquid phase from the stoichiometric
value, \(\Delta x_{\mathrm{Al}}^{\mathrm{liq}} = x_{\mathrm{Al}}^{\mathrm{liq}}-0.5\),
is summarized in Fig.~\ref{fig:delta_x}. At all pressures considered here, the
magnitude of this deviation does not exceed 0.005. Thus, within the accuracy
of the present simulations, no systematic enrichment of the liquid phase by
either component is observed. This result provides direct atomistic evidence
that melting of B2-NiAl remains congruent in the investigated pressure range.

\begin{figure}[h]
	\centering
	\includegraphics[width=0.95\columnwidth]{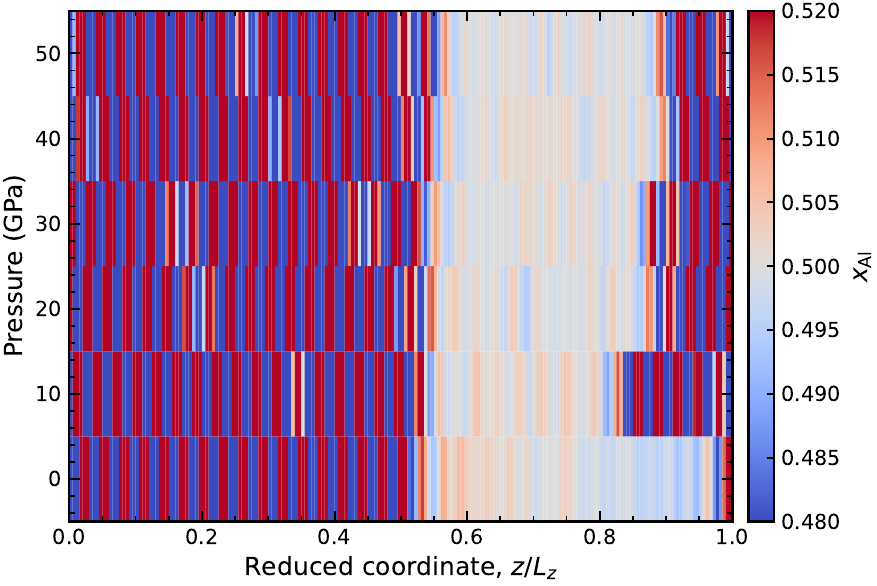}
	\caption{\label{fig:cmap_view}
		Two-dimensional composition map summarizing the spatial distribution
		of aluminum across the coexistence cell for all investigated
		pressures. Red denotes Al-rich layers, blue denotes Ni-rich layers,
		and the nearly white region corresponds to the liquid phase, where
		the composition is close to equiatomic. The alternating red/blue
		pattern in the solid part reflects the chemical ordering of the B2
		phase, whereas the broad pale region confirms the presence of an
		extended liquid slab with no noticeable chemical segregation.
	}
\end{figure}

For completeness, Fig.~\ref{fig:cmap_view} presents a compact two-dimensional
composition map for all studied pressures. In this representation, red
corresponds to Al-rich layers, blue to Ni-rich layers, and the nearly white
region corresponds to the liquid part of the coexistence cell, where the local
composition is close to equiatomic. In the solid region, the alternating
red/blue pattern reflects the ordered nature of the B2 phase, while the
roughly half-cell-wide pale region highlights the persistence of an extended,
chemically nearly uniform liquid slab. Although this representation is mainly
illustrative, it provides a convenient visual summary of the absence of any
pronounced pressure-driven chemical segregation.

\subsection{\label{sec:structure}Crystallization of the undercooled liquid: PTM and B2-order analysis}

To further characterize the structural response of the liquid upon cooling,
we performed melt--quench simulations at selected pressures and cooling
rates. Starting from equilibrated liquid configurations, the systems were
cooled to 300~K at fixed pressure and the resulting structures were analyzed
using polyhedral template matching (PTM) together with the local B2
chemical-order parameter introduced in Sec.~\ref{sec:ptm_method}. The final
state is expected to depend strongly on the cooling rate, as demonstrated,
for example, for the Cantor alloy, where rapid cooling promotes
amorphization whereas slower cooling favors crystallization
\cite{zhou2025amorphization}. A qualitatively similar cooling-rate dependence
is observed here for B2-NiAl.

For each cooling simulation, the final configuration at 300~K was analyzed
using PTM with an RMSD cutoff of 0.1. Each atom was assigned to one of the
mutually exclusive BCC, FCC, HCP, ICO, or \textit{Other} structural classes,
and the corresponding fractions were calculated relative to the total number
of atoms. These fractions characterize the local geometrical structure of
the quenched state, whereas chemical B2 ordering was analyzed separately
within the BCC-like population using the order parameter defined above.

\begin{table}[h]
	\centering
	\caption{Final PTM fractions (\%) obtained after quenching B2-NiAl from
		the liquid state to 300~K at different pressures and cooling rates.}
	\label{tab:ptm_fractions_compact}
	\begin{tabular}{ccccccc}
		\hline
		Pressure & Cooling rate & Other & \textbf{BCC} & ICO & HCP & FCC \\
		(GPa) & (K/ps) & (\%) & (\%) & (\%) & (\%) & (\%) \\
		\hline
		0  & 27.0 & 99.6 & 0.2  & 0.2 & 0.0 & 0.0 \\
		0  & 5.4  & 97.8 & 1.8  & 0.2 & 0.1 & 0.1 \\
		0  & 1.0  & 86.1 & 13.5 & 0.1 & 0.2 & 0.1 \\
		0  & \textbf{0.5} & \textbf{69.7} & \textbf{29.6} & 0.0 & 0.2 & 0.4 \\
		\hline
		30 & 33.0 & 99.1 & 0.3  & 0.3 & 0.2 & 0.1 \\
		30 & 6.6  & 97.9 & 1.3  & 0.3 & 0.3 & 0.2 \\
		30 & 1.0  & 45.3 & 53.5 & 0.1 & 0.7 & 0.4 \\
		30 & \textbf{0.5} & \textbf{25.3} & \textbf{63.4} & 0.0 & 0.9 & 0.3 \\
		\hline
	\end{tabular}
\end{table}

The results in Table~\ref{tab:ptm_fractions_compact} show that the structural
evolution is dominated by the development of a BCC-like crystalline
framework. At the highest cooling rates, crystallization is strongly
suppressed: almost all atoms are classified as \textit{Other}, while the
crystalline fractions remain negligible, consistent with an effectively
vitrified state. As the cooling rate is reduced, the BCC fraction increases
markedly. This trend is substantially stronger at 30~GPa, where the BCC
fraction reaches 63.4\% at 0.5~K/ps, compared with 29.6\% at the same
cooling rate at 0~GPa. Thus, both slower cooling and compression promote
crystallization of the undercooled liquid into a bcc-derived structure.

Small HCP- and FCC-like populations also appear during crystallization, but
remain minor relative to the BCC-like fraction. These environments are
therefore interpreted as locally distorted or defect-associated structures
rather than as competing bulk crystalline phases. The ICO fraction remains
negligible throughout the investigated cooling-rate range.

\begin{figure}[h]
	\centering
	\includegraphics[width=0.95\columnwidth]{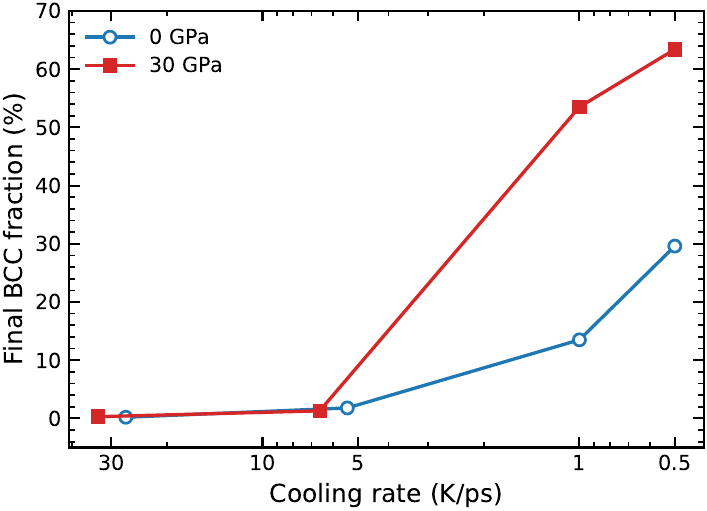}
	\caption{\label{fig:final_bcc_count}
		Final fraction of BCC-like atoms as a function of cooling rate for
		0 and 30~GPa. Slower cooling promotes crystallization at both
		pressures, with a substantially stronger increase of the BCC
		fraction at 30~GPa.
	}
\end{figure}

Figure~\ref{fig:final_bcc_count} emphasizes the strong cooling-rate
dependence of the final BCC fraction and the enhanced tendency toward
crystallization under compression. The systematic increase of the BCC
population with increasing available relaxation time indicates that the
structural outcome is controlled by the competition between the cooling
timescale and the kinetics of crystal formation.

\begin{figure}[h]
	\centering
	\includegraphics[width=0.95\columnwidth]{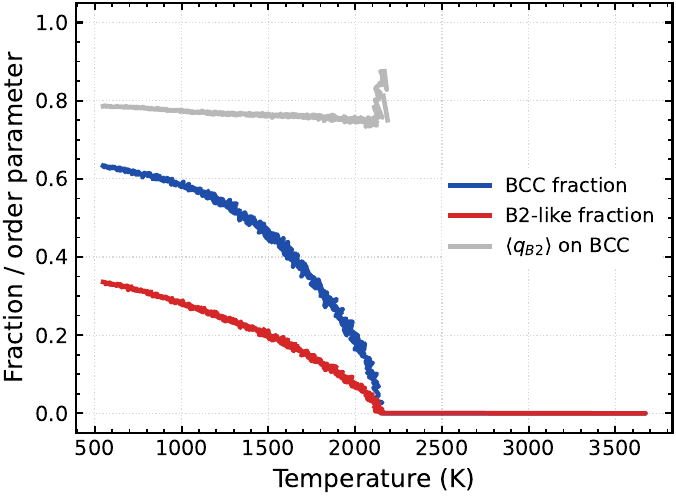}
	\caption{\label{fig:b2_fraction}
		Temperature evolution of the BCC-like fraction, the B2-like
		fraction, and the average B2 order parameter evaluated over BCC-like
		atoms during quenching at 30~GPa with a cooling rate of 0.5~K/ps.
		The rise of the BCC fraction marks crystallization, whereas the
		subsequent increase in chemical order indicates progressive
		development of the B2 arrangement.
	}
\end{figure}

Additional insight into the crystallization mechanism is provided by the
temperature evolution of the BCC and B2-like fractions shown in
Fig.~\ref{fig:b2_fraction}. At high temperatures, both quantities remain
close to zero, consistent with a structurally disordered liquid. Upon
cooling, the BCC fraction rises sharply, indicating the formation of a
bcc-like crystalline framework. The B2-like fraction remains lower than the
total BCC fraction, while the average B2 order parameter evaluated over
BCC-like atoms increases more gradually. This indicates that the formation
of the bcc-like geometry precedes the establishment of full chemical B2
ordering: crystallization first produces a bcc-derived structural framework,
followed by progressive ordering of the Al and Ni sublattices.

\begin{figure}[h]
	\centering
	\includegraphics[width=0.95\columnwidth]{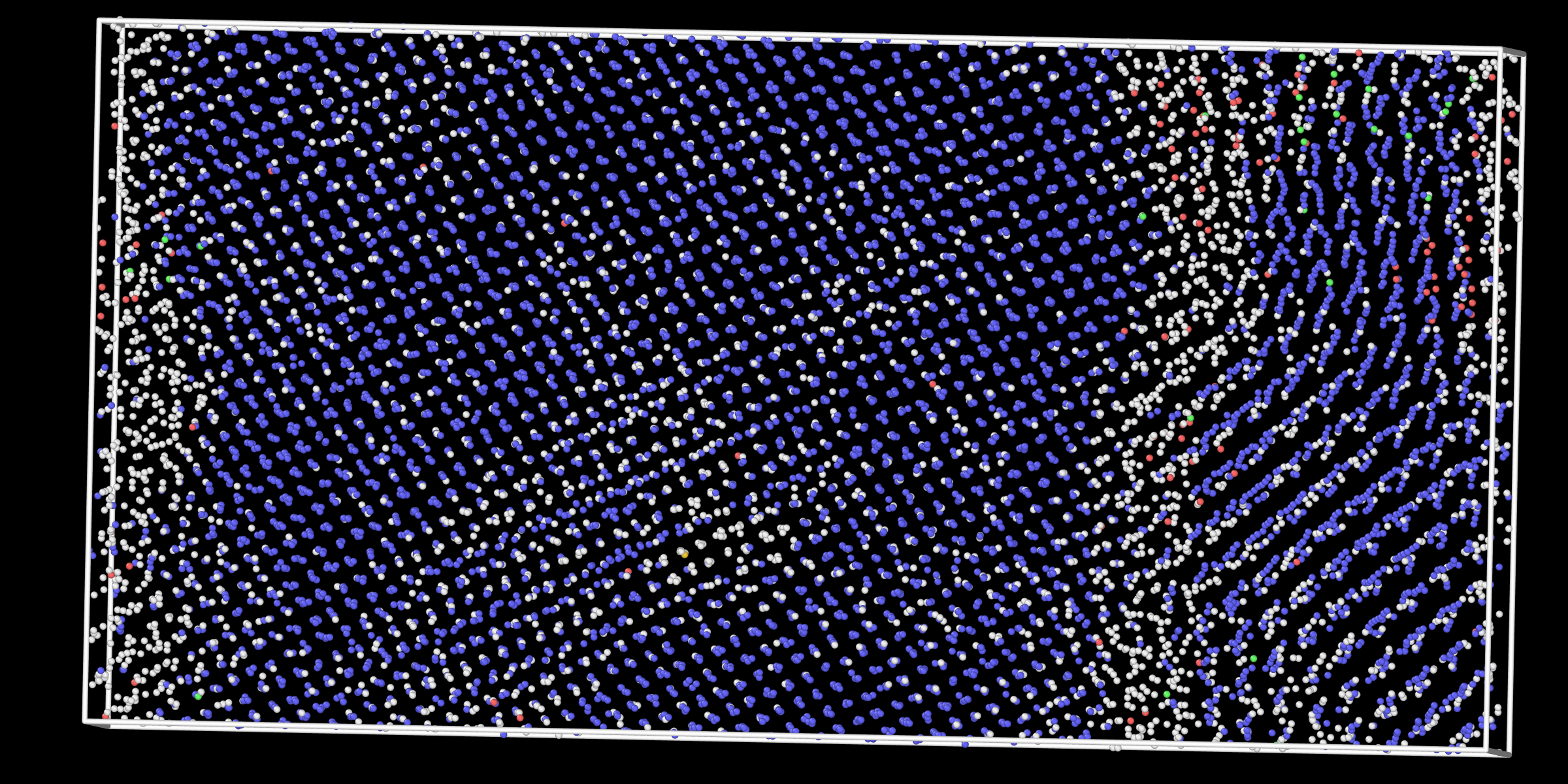}
	\caption{\label{fig:polycrystal}
		Final quenched configuration obtained at 30~GPa and a cooling rate
		of 0.5~K/ps, colored by PTM structure type with an RMSD cutoff of
		0.1. White denotes atoms classified as Other, blue denotes BCC-like
		atoms, red denotes HCP-like atoms, and green denotes FCC-like atoms.
		The final state is polycrystalline, with several crystallites of
		different orientations.
	}
\end{figure}

A representative final configuration obtained at 30~GPa and a cooling rate
of 0.5~K/ps is shown in Fig.~\ref{fig:polycrystal}. The final state is
polycrystalline, with several crystallites having different orientations.
Most atoms in the crystalline regions are classified as BCC-like, whereas
the small HCP-like population is concentrated mainly in intergranular and
locally distorted regions. No appreciable icosahedral population is observed.

An additional spatial analysis shows that BCC-like atoms that do not satisfy
the B2 chemical-order criterion are not preferentially localized at grain
boundaries. Instead, they are distributed throughout the crystallites. This
observation further supports the interpretation that the bcc-like structural
framework develops before complete chemical B2 ordering is established.

Taken together, the melt--quench simulations show that the undercooled
B2-NiAl liquid crystallizes more readily at lower cooling rates and at higher
pressure. The resulting crystalline state is predominantly bcc-derived,
while chemical B2 ordering develops progressively within this framework.
Although these simulations do not constitute a direct test of congruent
melting, they provide complementary evidence that crystallization retains a
strong preference for B2-related ordering and does not produce chemically or
structurally distinct competing crystalline products.


\section{\label{sec:conclusion}Conclusion}

We have studied melting of ordered B2-NiAl under pressure using
neural-network molecular dynamics with a Deep Potential interatomic model. The melting curve was determined from two-phase solid--liquid coexistence simulations over a broad pressure range and was found to increase monotonically with pressure. Independent single-phase calculations of the enthalpy and volume changes upon melting show that the resulting Clapeyron slope is consistent with the derivative of the fitted melting curve, providing an internal thermodynamic check of the coexistence results. A layer-resolved composition analysis further shows that the coexisting liquid remains essentially equiatomic at all investigated pressures, which provides direct atomistic evidence that melting of B2-NiAl remains congruent within the present model. Complementary melt--quench simulations indicate that slower cooling and higher pressure promote crystallization into a bcc-derived polycrystalline state with gradually developing B2 chemical order. Taken together, these results provide a coherent atomistic description of pressure-driven melting in B2-NiAl and clarify both the melting line and the character of melting under compression.

\vspace{1em}

\section*{Supplementary Material}

The supplementary material contains the trained Deep Potential model 
for all molecular-dynamics simulations reported in this work. The model is provided in the serialized DeePMD format and can be used directly with the DeePMD-kit interface implemented in LAMMPS. It reproduces the interatomic potential employed for the calculations of the melting curve, thermodynamic properties, solid--liquid coexistence, and melt-quench simulations discussed in the manuscript. The accompanying supplementary file is intended to enable reproduction of the molecular-dynamics calculations and further simulations
of the Al--Ni system within the range of thermodynamic conditions considered in this study. 
	
\begin{acknowledgments}
	The data that supports the findings of this study are available from the corresponding author upon reasonable request. The authors acknowledge the JIHT RAS Supercomputer Centre, and the Shared Resource Centre ``Far Eastern Computing Resource'' IACP FEB RAS for providing computing time.
\end{acknowledgments}

		\appendix
		


%

	\end{document}